\documentclass[letterpaper,10pt]{article} 

\usepackage{osameet3}
\usepackage{amsmath}
\usepackage{amssymb}
\usepackage{xcolor}
\usepackage{caption}
\usepackage{subcaption}
\usepackage{comment}

\colorlet{linkcolor}{teal!65!black}
\usepackage[colorlinks=true, allcolors=linkcolor]{hyperref}

\graphicspath{{./figs/}}

\begin{document}
\advance\voffset by -1.5em 
\title{Channel Matching: An Adaptive Technique to Increase the Accuracy of Soft Decisions}

\author{Reza Rafie Borujeny and Frank R. Kschischang}
\address{Department of Electrical and Computer Engineering\\
University of Toronto, Toronto, Ontario, M5S 3G4, Canada.}
\email{\href{mailto:rrafie@ece.utoronto.ca}{rrafie@ece.utoronto.ca}, \href{mailto:frank@ece.utoronto.ca}{frank@ece.utoronto.ca}}

\copyrightyear{2021}
\vspace{-1.3em}
\begin{abstract} 
Nonlinear interference is modeled by a time-varying conditionally Gaussian
channel.  It is shown that approximating this channel with a time-invariant
channel imposes considerable loss in the performance of channel decoding.
An adaptive method to maintain decoding performance is described.
\end{abstract}
\vspace{-0.1em}
\section{Introduction}
To meet the stringent reliability requirements of optical networks,
some form of forward-error-correction (FEC) is usually employed.
One important class of FEC schemes
uses an inner low-density parity-check
(LDPC) coded modulation scheme---which may suffer from an error
floor---concatenated with an algebraic outer code. Two popular approaches
used to achieve the required bit error rate (BER) seen by the outer code
are bit-interleaved coded modulation (BICM) \cite{caire1998bit} and
multi-level coding (MLC)\cite{wachsmann1999multilevel}. 

It is well-known that the fiber nonlinearity causes inter-channel
interactions that, eventually, cap the spectral efficiency of a
wavelength-division multiplexed (WDM) system \cite{essiambre10jlt}. Each
WDM channel leaks energy into neighboring channels which is impossible to
perfectly mitigate due to the absence of the neighboring WDM channels at a
given receiver. The resulting inter-channel interference is usually
considered as a nuisance and is referred to as nonlinear interference noise
(NLIN) \cite{dar2014accumulation}.  Usually, NLIN is assumed to be
\emph{additive} and is modeled by a Gaussian random vector whose covariance
is a function of the average power of the neighboring channels, the
instantaneous power of the channel of interest, the modulation and
demodulation format and the fiber parameters. Amongst other factors, the
average power of the neighboring channels has the most significant
contribution in the NLIN covariance \cite{poggiolini2012gn}. At the same
time, it is known that the average power of WDM channels can fluctuate
\cite{krummrich2004extremely,kilper2006transient} which, in turn, causes
fluctuations in the covariance of NLIN. If the fluctuations of the noise
parameters are not properly taken into account, the soft information fed
into the FEC decoder may be inaccurate. In this work, we study the
importance of having accurate soft information on the performance of the
decoder.

We take the inner MLC-based coded modulation scheme of
\cite{barakatain2020performance} for our baseline system, although similar
results are expected with BICM. We assume a time-varying conditionally
Gaussian channel in which, conditional on the input, the noise covariance
is a function of the average power of the WDM channels. We model the
fluctuations of the average power through the second order statistics of
the noise and consider the performance of the inner MLC scheme with two
main strategies:  with a \emph{fixed} estimate of the NLIN power and with
an \emph{adaptive} estimate of the NLIN power.  Our main finding is
that if one does not adaptively match the soft information to the
conditions of the channel, considerable degradations in decoding
performance are caused.  These results are also compared against a
hypothetical genie-aided decoder having access to perfect channel state
information.

It is shown that the simple strategy of re-estimating the noise covariance
based on the currently decoded codeword can significantly improve the
performance of the inner code. This simple idea is the main ingredient of
turbo equalization \cite{douillard1995iterative} and has been previously
used to compensate for channels with inter-symbol interference. This includes the compensation of intra-channel
effects in optical fiber (see \cite{djordjevic2008mitigation} and
references therein) as well as inter-channel nonlinear equalization
\cite{golani2019nlin}. The application of turbo equalization for channel
estimation has also been considered for some linear channels
\cite{song2004soft}.

\section{Simulation Setup and Channel Model}

\begin{figure}[t]
\centering
\begin{subfigure}[b]{0.6\textwidth}
  \centering
  \includegraphics[width=\textwidth]{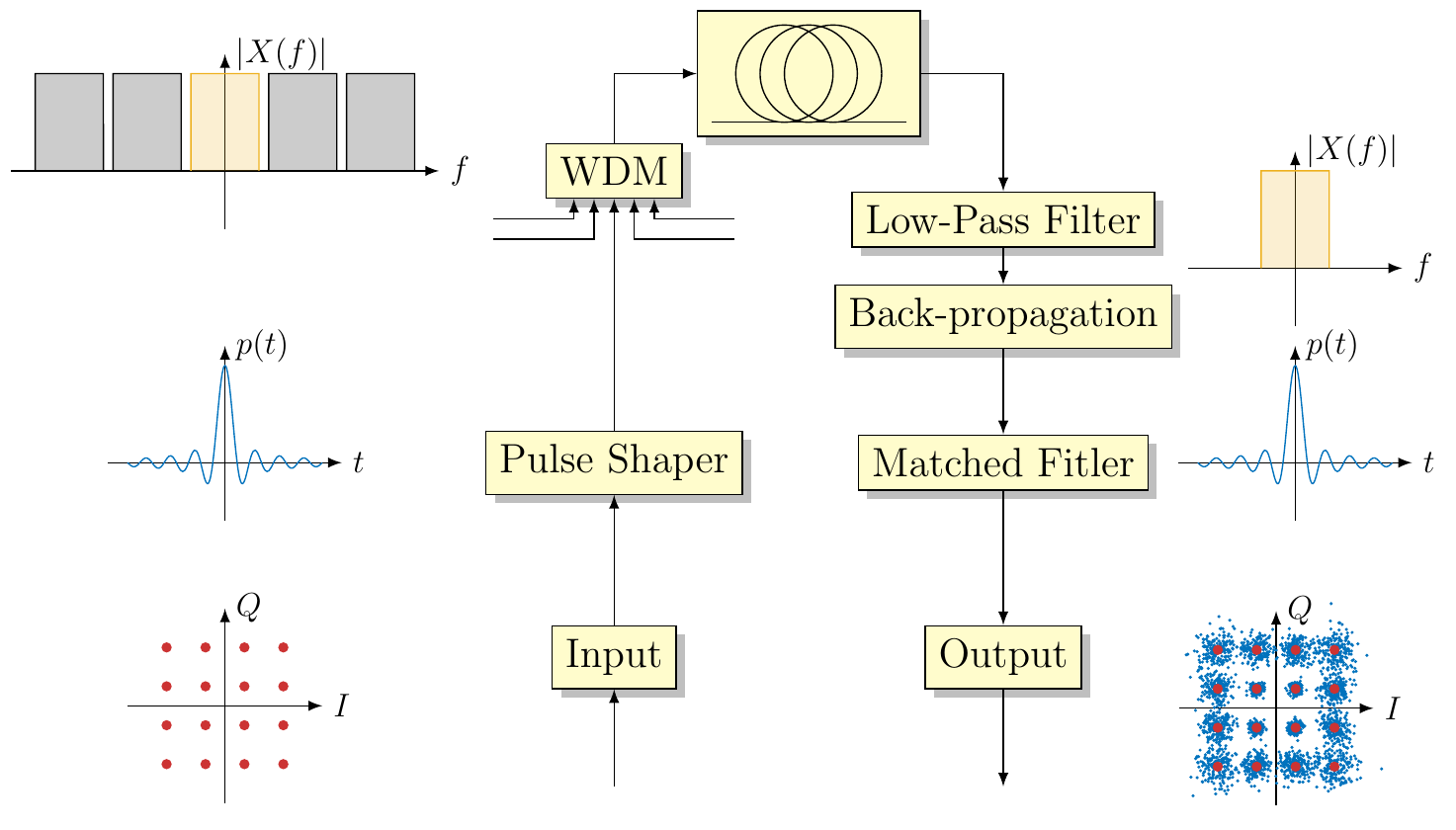}
  \caption{The channel model}
  \vspace{-1.5em}
  \label{fig:cm}
\end{subfigure}
\hfill
\begin{subfigure}[b]{0.3\textwidth}
  \centering
  \includegraphics[width=\textwidth]{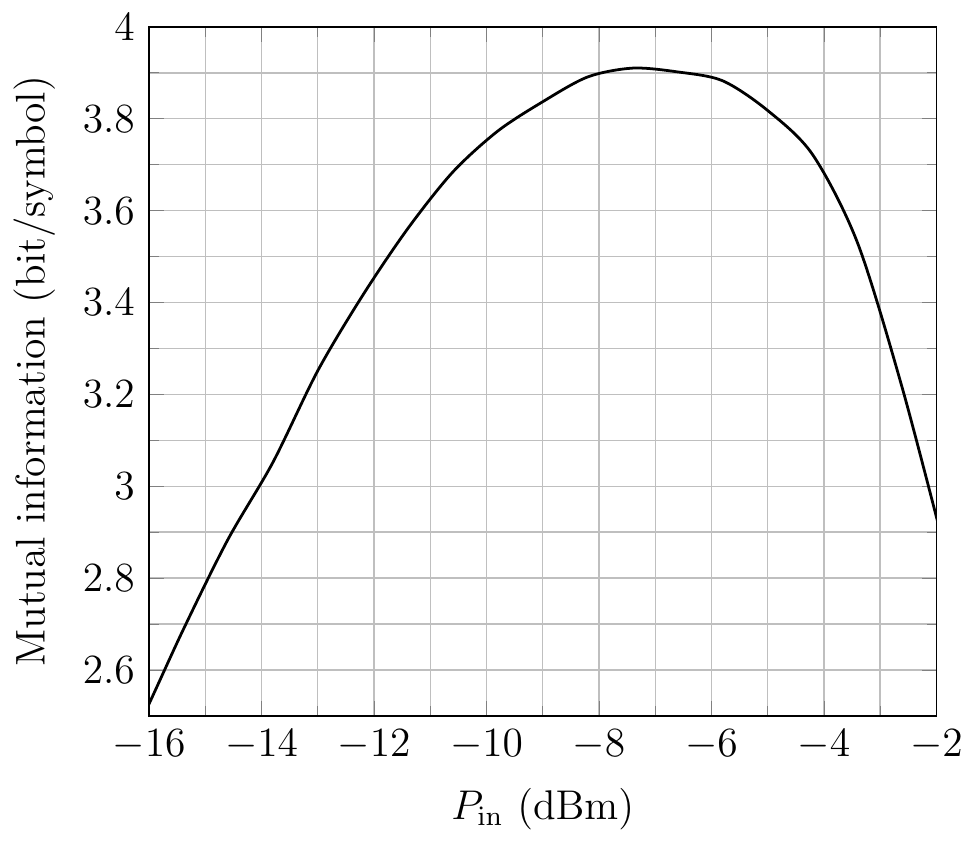}
  \caption{Mutual information}
  \vspace{-1.5em}
  \label{fig:mi}
\end{subfigure}
\caption{The blocks in the considered channel model are shown in (a). The mutual information achieved with uniform input distribution for the channel of interest in the simulated WDM system is illustrated as a function of the average input power (b).} 
\vspace{-1.5em}
\label{fig:mi_ber}
\end{figure}

We simulate a single-polarization WDM system with five channels over a
fiber of length 4500 km with the adaptive split-step Fourier method of
\cite{sinkin2003optimization}. In each run, the split-step model with
periodic boundary conditions is solved for trains of 3600 symbols. It is
assumed that each span of length $L_A = 50$ km is followed by an
erbium-doped fiber amplifier (EDFA). The fiber loss is set to $\alpha =
0.2~\text{dB}\cdot\text{km}^{-1}$. The nonlinearity coefficient is set to
$\gamma = 1.27~\text{W}^{-1}\cdot\text{km}^{-1}$ and the chromatic
dispersion coefficient is set to $\beta_2 =
-21.67\times10^{-24}~\text{s}^2\cdot\text{km}^{-1}$. The amplified
spontaneous emission (ASE) noise is assumed to be a circularly symmetric
white Gaussian process with power spectral density
$N_{\text{ASE}}^{\text{EDFA}} = \left(e^{\alpha L_A} - 1\right)h\nu
n_{\text{sp}} $ where $h=6.626\times 10^{-34}~\text{J}\cdot\text{s}$ is
Planck's constant, $\nu = 193.41~\text{THz}$ is the center frequency, and
$n_{\text{sp}} = 1$ is the spontaneous emission factor.  The channel
spacing is assumed to be 50~GHz and a 16-QAM constellation is used. About
6.67\% of the channel spacing is reserved for a guard band.
Root-raised-cosine pulses with 6.67\% excess bandwidth are used for pulse
shaping. As a result, the symbol rate per WDM channel is
$43.95~\text{Gsymbol}\cdot \text{s}^{-1}$. At the receiver, the channel of
interest---the center channel---is filtered and digitally back-propagated,
and the result is passed through a matched filter. The obtained points are commonly back rotated to undo the effect of cross-phase modulation. All WDM channels are
assumed to have the same average input power $P_{\text{in}}$. The achieved
mutual information under uniform input distribution is shown in Fig.
\ref{fig:mi}. With a maximum at $-6.8~\text{dBm}$, this
curve suggests that $P_{\text{in}}$ must be set to the optimal value of
$-6.8~\text{dBm}$.

We model data transmission in a WDM system with a time-varying
conditionally Gaussian channel with input random variable $X$ and output
random variable $Y$. The input alphabet $\mathcal{X}$ is the 16-QAM
constellation in use, while the output alphabet $\mathcal{Y}$ is the whole
complex plane (see Fig. \ref{fig:cm}). 

Ideally, one wishes to operate on the optimal input power, or very close to
it, so that the data rate can be maximized with aid of an appropriate
choice of FEC.  Because of the dynamics of the network, some power
fluctuations may happen which in turn result in reductions in the
achievable information rates. Our goal here is to capture such fluctuations
and make sure that the FEC block is aware of them. This is achieved by
re-estimating the noise parameters and making sure that the soft information
used by the FEC decoder is accurate.

\section{Multi-level Coding}

We consider a similar MLC transmission scheme as in
\cite{barakatain2020performance} with a 16-QAM constellation and an inner
LDPC code of rate $0.63$ which gives an overall inner coded modulation
information rate of 3.63 bits per symbol.  We study the performance of the
inner coded modulation scheme under three scenarios, namely, obtaining
noise covariance based on the optimal input power, a genie-aided decoder
and an iterative channel estimation. We define the \emph{survivability} of
the inner coded modulation as the range of power fluctuations that it can
tolerate while still maintaining the BER below a prescribed target BER. In
our running example, the target BER is set to $10^{-3}$. We compare these
three decoding methods in terms of their survivability.

\begin{figure}[h]
\centering
\begin{subfigure}[b]{0.5\textwidth}
  \centering
  \includegraphics[width=\textwidth]{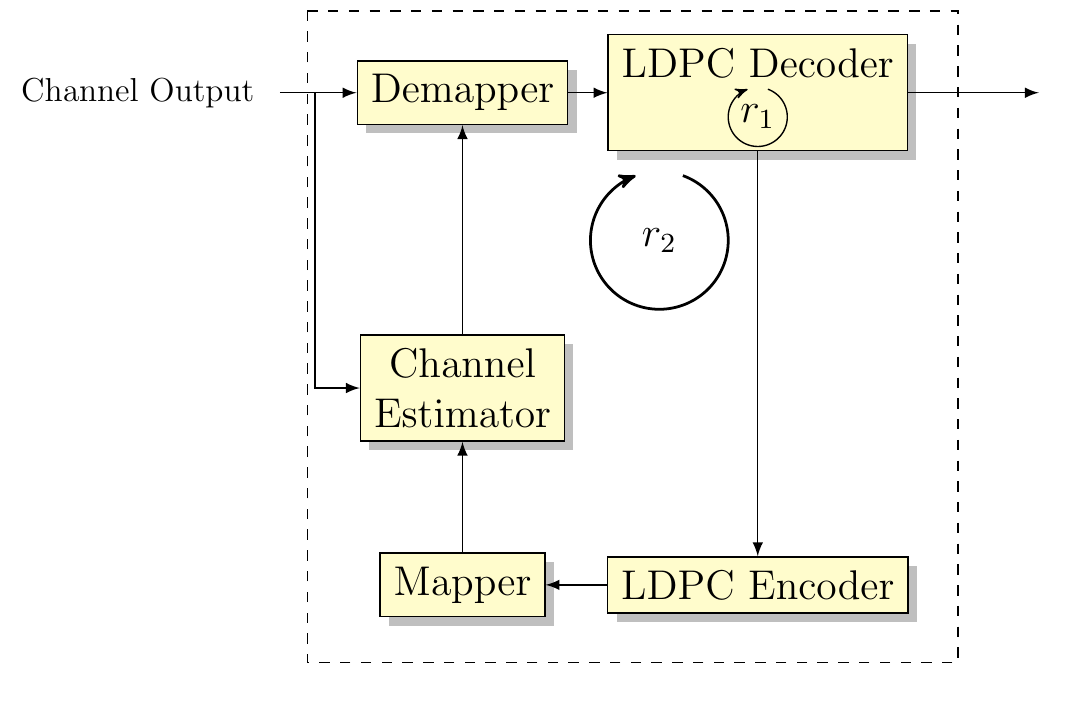}
  \caption{The decoder}
  \vspace{-1.5em}
  \label{fig:decoder}
\end{subfigure}
\hfill
\begin{subfigure}[b]{0.4\textwidth}
  \centering
  \includegraphics[width=\textwidth]{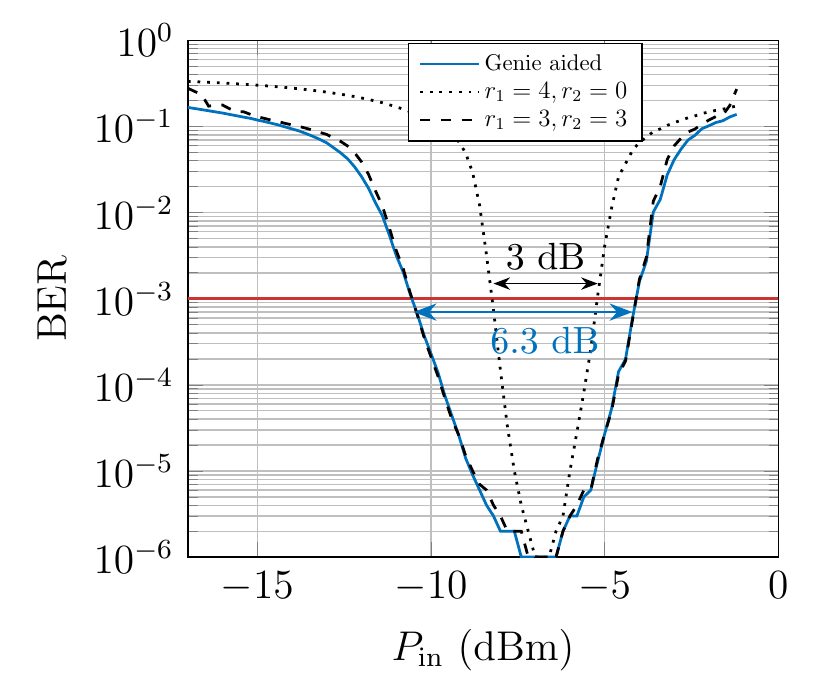}
  \caption{BER vs. average input power}
  \vspace{-1.5em}
  \label{fig:ber}
\end{subfigure}
\caption{Structure of the adaptive decoder of this work is shown in (a). Performance of the inner coded modulation scheme, in terms of BER, is shown in (b).} 
\vspace{-1.5em}
\label{fig:decoder_ber}
\end{figure}
\noindent\textbf{Fixed Noise Covariance:} In this scenario, the inner decoder
calculates the soft information based on the optimal input power, i.e., the
noise parameters are those induced by the optimal input power.  This means
that the noise covariance is assumed to be fixed, despite the fact that the
actual average power experiences fluctuations. The results are shown in
Fig.~\ref{fig:ber}. When the receiver uses a fixed set of noise parameters
based on the optimal power with a target BER of $10^{-3}$, the MLC is
successful as long as the average input power (in dBm) lies in $[-8.2,
-5.2]$. That is, the survivability of this method is 3~dB. 

\noindent\textbf{Perfect Channel State Information:} In this scenario, it is assumed
that a genie provides the inner decoder with the true noise statistics.
This corresponds to the availability of perfect channel state information
at the receiver. The corresponding power interval that the MLC performs
below the target BER of $10^{-3}$ is $[-10.5, -4.2]$ dBm with a
survivability of 6.3~dB, i.e., about 3.3 dB more survivability in
comparison with having fixed estimates for the noise statistics. The
results are shown in Fig.~\ref{fig:ber}.

\noindent\textbf{Channel Matching:} In this scenario, the inner code initially
calculates soft information based on the optimal power. The decoded
codeword is then remapped to the corresponding constellation points
similarly to the way that the transmitter performs the mapping of the
codeword to the constellation points. By considering the difference between
these symbols and the actually received symbols, maximum likelihood
estimates of the noise statistics are obtained. The updated noise
statistics are then used to update the soft information fed to the decoder
in a turbo-equalizer fashion.  Fig.~\ref{fig:decoder} shows the structure of
the turbo LDPC decoder used in this work. There are two parameters, namely
$r_1$ and $r_2$, that determine the number of iterations performed: $r_1$
counts the number of iterations performed by the LDPC decoder before
re-estimating the noise parameters while $r_2$ counts the number of
``turbo'' iterations performed to re-estimate the noise parameters. The
performance of this method is illustrated in Fig. \ref{fig:ber}.  Depending
on the choice of $r_1$ and $r_2$, different survivabilities can be
obtained. This provides an interesting trade-off between decoding
complexity and reliability. One can see that with 3 turbo iterations and 3
LDPC decoding iterations, the survivability is virtually identical to that
of the genie-aided decoder.

\bibliographystyle{osajnl}
\bibliography{mybib}
\end{document}